\documentclass[dvips]{article}
\usepackage{amsmath,amssymb}
\usepackage{icrctc07}
\title{Diffuse Gamma-Rays Produced in Cosmic-Ray Interactions and the TeV-band Spectrum of RX J1713.7-3946}
\shorttitle{Diffuse Gamma-Rays in Cosmic-Ray Interactions}
\authors{C.-Y. Huang, M. Pohl, S.-E. Park and C. D. Daniels}
\shortauthors{C.-Y. Huang et al.}
\afiliations{Department of Physics and Astronomy, Iowa State University, Ames, IA 50011}
\email{huangc@iastate.edu}

\abstract{We employ the Monte Carlo particle collision code DPMJET3.04 to determine the multiplicity spectra of various secondary particles (in addition to $\pi^0$'s) with $\gamma$'s as 
the final decay state, that are produced in cosmic-ray ($p$'s and $\alpha$'s) interactions with the interstellar medium. We derive an easy-to-use $\gamma$-ray production matrix for cosmic rays 
with energies up to about 10 PeV. This $\gamma$-ray production matrix is applied to the GeV excess in diffuse Galactic $\gamma$-rays
observed by EGRET, and we conclude the non-$\pi^0$ decay components are insufficient to explain the GeV excess, although they have contributed a different spectrum from 
the $\pi^0$-decay component. We also test the hypothesis that the TeV-band $\gamma$-ray emission of the shell-type SNR RX J1713.7-3946 
observed with HESS is caused by hadronic cosmic rays which are accelerated by a cosmic-ray modified shock. By the $\chi^2$ statistics, we find a continuously softening 
spectrum is strongly preferred, in contrast to expectations. A hardening spectrum has about 1\% probability to explain the HESS data, but then only if a hard 
cutoff at 50-100 TeV is imposed on the particle spectrum.}

\begin{document}
\maketitle

\subsection{The $\gamma$-Ray Production Matrix}
It is generally believed that diffuse galactic $\gamma$-rays are a good probe of 
the production sites and also the propagation of accelerated charged particles in the 
Galactic plane (see \cite{Erlykin98} and references therein). Exact knowledge of the $\gamma$-ray source spectrum resulting from hadronic interactions is 
of prime importance for a proper physical interpretation for the diffuse $\gamma$-rays. The production of neutral pions 
and their subsequent decay to $\gamma$-rays is thought to be the 
principal mechanism for the $\gamma$-ray hadronic component in cosmic ray
interactions, and a number of parameterizations have been developed to describe $\gamma$-ray production in $pp$-collision \cite{Kamae06,Kelner06}. 
On the other hand, in addition to the neutral pion production, the direct $\gamma$-ray production in cosmic-ray interaction was found important \cite{Huang07}, as was
the production of $(\Sigma^\pm,~\Sigma^0)$, $(K^\pm,~K^0)$, and $\eta$ particles, all of which eventually decay 
into $\gamma$-rays \cite{Kelner06}. Also, the helium nuclei in cosmic rays and the interstellar medium
are expected to contribute about 20-30\% to the secondary production in cosmic-ray interactions, for which the
multiplicity spectra may be different. In a hadronic interaction, the multiplicity of non-$\pi^0$ secondaries with $\gamma$-rays as the final decay states is about 50\% of 
the $\pi^0$ multiplicity. These non-$\pi^0$ secondaries in hadronic interactions might present a different $\gamma$-ray spectral profile.

This work presents a careful study of the $\gamma$-ray production in cosmic-ray (both $p$ and $\alpha$) interactions, by accounting for all decay processes including 
the direct production. For that purpose we employ the 
event generator DPMJET-III \cite{Roesler00} to simulate secondary productions in both p-generated and 
He-generated interactions. We include all relevant secondary particles with $\gamma$-rays as the 
final decay products. For the composition of the ISM, we assume 90\% protons, 10\% helium nuclei, 0.02\% carbon, and 0.04\% oxygen. 
Around the energy of $\pi^0$ production threshold, where DPMJET appears unreliable, we apply a parametric model \cite{Kamae06}, 
that includes the resonance production for the $\pi$ production. We thus derive a $\gamma$-ray production matrix for cosmic rays 
with energies up to about 10 PeV that can be easily used to interpret the spectra of cosmic $\gamma$-ray sources. 

We consider all the decay channels and their decay fractions published by the Particle Data Group to account for all secondaries (resonances included) calculated 
by DPMJET and the parametric method. The calculation shows the non-$\pi^0$ resources in hadronic interactions have contributed about 20\% of the total $\gamma$-ray 
photons, mostly from directly produced $\gamma$-rays and decays of $\eta$, $K^0_L$ and $K^0_S$ \cite{Huang07}.

In the cosmic-ray interactions, we calculate the $\gamma$-ray spectrum density contributed by decays of unstable secondary particles
\begin{eqnarray}\label{Eq:gaSpectra}
\frac{Q_{\gamma}(E_\gamma)}{n_{ISM}}  = \sum_k \int dE N_{CR}(E) c\beta\, \sigma (E)\, \frac{dn_{k,\gamma}}{dE_\gamma}
\end{eqnarray}
where $\frac{dn_{k,\gamma}}{dE_\gamma}=\frac{dn_{k,\gamma}}{dE_\gamma}(E_\gamma,E)$ is the $\gamma$-ray decay spectrum from secondary species $k$. Eq.~(\ref{Eq:gaSpectra}) 
can be re-written into
\begin{eqnarray} \label{Eq:ProductionMatrix}
& & \frac{Q_{\gamma}(E_i)}{n_{ISM}}\nonumber\\
&=& \sum_{j} \Delta E_{j}\ N_{CR}(E_j) c\beta_j \sigma (E_{j}) \sum_{k} \frac{dn_{k,\gamma}}{dE_\gamma}\nonumber\\
&=& \sum_j \Delta E_{j} N_{CR}(E_j) c\beta_{j} \sigma_j  \mathcal{M}_{ij}
\end{eqnarray}
thus reducing this problem to a matrix operation with the $\gamma$-ray production matrix $\mathcal{M}_{ij}$ for which, each element 
$\mathcal{M}_{ij}=\sum_{k} \frac{dn_{k,\gamma}}{dE_\gamma}$ shows the value of 
the resultant particle energy spectrum $\frac{dn}{dE}|_{E_{\gamma}=E_i,E_{CR}=E_j}$, with $j$ being the index for the generating cosmic-ray particle ($p$ or $\alpha$) 
and $i$ being the index indicating the $\gamma$-ray energy. The energy binnings $E_i$ and $E_j$ are defined with good resolutions \cite{Huang07}. 

We then use 
the $\gamma$-ray production matrix to analyze the observed spectra of diffuse Galactic emission and of the shell-type SNR RX J1713.7-3946. 
\subsection{The GeV Excess}
\begin{figure}[h]
\vspace*{-0.6cm}
\begin{center}
\includegraphics[width=0.48\textwidth]{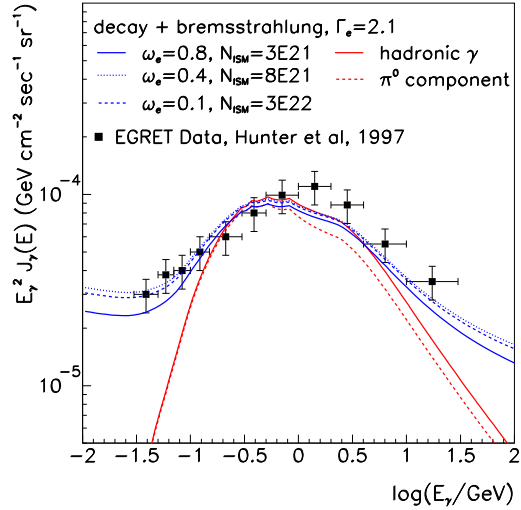}
\end{center}
\vspace*{-0.7cm}
\caption{Diffuse $\gamma$-ray spectrum at GeV range, shown in comparison with data for the inner part of Galaxy at $315^0\le l \le 345^0$ and $|b|\le 5^0$\cite{Hunter97}.
See text for discussions.}\label{Fig:GeVBump}
\end{figure}
With the $\gamma$-ray production matrix, we calculate the diffuse $\gamma$-ray spectrum generated by the observed cosmic-ray spectrum \cite{Mori97}. 
Fig.~\ref{Fig:GeVBump} shows the observed GeV-band $\gamma$-ray emission from the inner Galaxy \cite{Hunter97} in comparison with the contributions 
from $\pi^0$ decay as well as bremsstrahlung emission describe by a power-law spectrum $\Phi_{\textrm{B}}(E)$:
\begin{eqnarray}\label{EQ:BremssPowerLaw}
\Phi_{\textrm{B}}(E) & \simeq & 1.3 \times10^{-8} \frac{\omega_e}{0.1~\textrm{eV/cm}^3}
                                    \cdot \frac{N_{ISM}}{10^{22}~\textrm{cm}^{-2}} \nonumber \\
                     & \times &     \left(\frac{E}{100~\textrm{MeV}}\right)^{2.0-\Gamma_e} 
                                    \frac{\textrm{erg}}{\textrm{cm}^2~\textrm{sec}~\textrm{sr}}
\end{eqnarray}
with power-law spectral index $\Gamma_e=2.1$, the electron energy density 
$\omega_e=0.1,~0.4,~0.8~\textrm{eV/cm}^3$, and gas column density 
$N_{ISM}=3\cdot 10^{22}$, $8\cdot 10^{21}$, $3\cdot 10^{21}\textrm{cm}^{-2}$ respectively. 
The generating cosmic-rays are assumed with an energy 
density $\rho_E=0.75~\textrm{eV/cm}^3$. 
Models based on the locally observed cosmic-ray spectra generally
predict a softer spectrum for the leptonic components, even after accounting for inverse 
Compton emission \cite{Hunter97}, so we may in fact overestimate the GeV-band intensity 
of the leptonic contribution.
Nevertheless, it is clearly seen in this figure, that in the total intensity an over-shooting around 
$E_{\gamma} \simeq ~\textrm{300-600~MeV}$ appears in the modelled $\gamma$-ray energy 
distribution, whereas a deficit is present above 1~GeV. The observed spectrum of diffuse emission is 
always harder than the model spectrum, and we therefore conclude that an inaccurate description of 
hadronic $\gamma$-rays is ruled out as the origin of the GeV excess.
\begin{figure}[t]
\vspace*{-0.8cm}
\begin{center}
\includegraphics[width=0.48\textwidth]{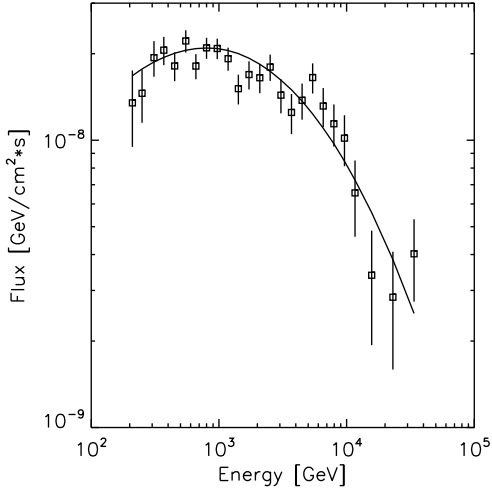}
\end{center}
\vspace*{-0.8cm}
\caption{The TeV-band $\gamma$-ray spectrum observed from RX~J1713.7-3946 with HESS \cite{Aharonian06AA}, 
shown in comparison with the best-fit model of hadronic $\gamma$-ray production of Eq.~(\ref{EQ:SNRSpectra}).}\label{Fig:SNRSpectra}
\end{figure}
\subsection{TeV-band Emission from RX~J1713.7-3946}
\begin{figure}[t]
\vspace*{-0.6cm}
\begin{center}
\includegraphics[width=0.48\textwidth]{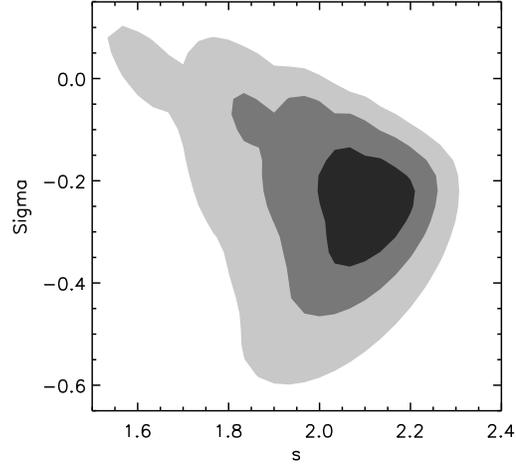}
\end{center}
\vspace*{-0.8cm}
\caption{The confidence regions for the spectral curvature $\sigma$ and the spectral index $s$. The shaded areas correspond to probabilities 68\% (1 sigma), 
95\% (2 sigma) and 99.7\% (3 sigma).}\label{Fig:CLSigmaS}
\end{figure}

\begin{figure}[h]
\vspace*{-0.6cm}
\begin{center}
\includegraphics[width=0.50\textwidth]{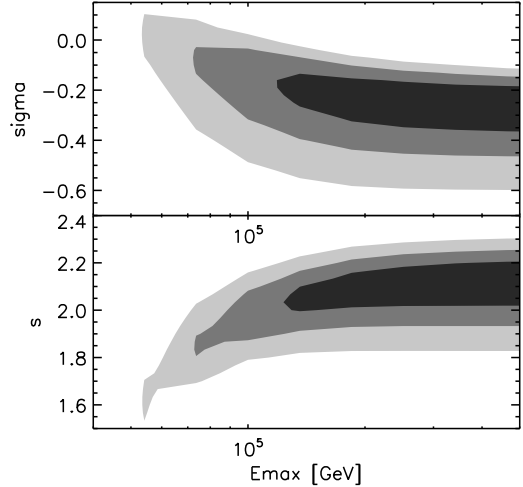}
\end{center}
\vspace*{-0.6cm}
\caption{Confidence contours as in Fig.~\ref{Fig:CLSigmaS}.}\label{Fig:CLSigmaSEmax}
\end{figure}

For the TeV-band $\gamma$-ray spectrum of the shell-type SNR J1713.7-3946 observed by the HESS collaboration \cite{Aharonian06AA}, we use the 
$\gamma$-ray production matrix to test cosmic-ray acceleration models \cite{Amato06,Berezhko99}, which predict a continuous hardening of the cosmic-ray spectrum up to a
high-energy cutoff. We therefore parametrize the spectrum of accelerated hadrons as
\begin{eqnarray}\label{EQ:SNRSpectra}
N(E)=N_0 \left({E\over {E_0}}\right)^{-s+\sigma \ln {E\over {E_0}}}
\Theta\left[E_{\rm max}-E\right]
\end{eqnarray}
where $\Theta$ is the step function and $E_0=15$~TeV is a normalization chosen to render variations in the power-law index $s$ statistically independent from the 
choice of spectral curvature, $\sigma$. The cutoff energy, $E_{\rm max}$, is a free parameter. The normalization $N_0$ is obtained by normalizing 
both the data and the model to the value at 0.97~TeV. By the $\chi^2$ statistics, we obtain the best-fitting values and the confidence
ranges of the three parameters, $E_{\rm max}$, $s$, and $\sigma$, given values as $s=2.13$, $\sigma=-0.25$, 
and $E_{\textrm{max}} \ge 200$~TeV, i.e., no cutoff. The best fit, shown in Fig.~\ref{Fig:SNRSpectra}, involves a continuous softening and is thus not 
commensurate with expectations based on acceleration at a cosmic-ray modified shock \cite{Amato06,Berezhko99}. Note the need for data in the energy 
range between 1~GeV and 200~GeV that may be provided by GLAST in the near future. Fig.~\ref{Fig:CLSigmaS} and Fig.~\ref{Fig:CLSigmaSEmax} shows the confidence ranges of 
the parameters in Eq.~(\ref{EQ:SNRSpectra}), for $\sigma$ vs. $s$, $\sigma$ vs. $E_{\textrm{max}}$ and $s$ vs. $E_{\textrm{max}}$ respectively, with confidence levels 
of 1, 2 and 3 sigma. The contour for $E_{\textrm{max}}$ are open toward higher energies, indicating that a cut-off is not statistically required. The analysis shows that 
a very high $E_{\textrm{max}}$ is possible, but requires a negative spectral curvature, i.e.,  $\sigma < 0$,  with confidence more than 95\%, in contrast to the expectation of 
standard cosmic-ray modified shock models. 

\begin{figure}[t]
\vspace*{-1cm}
\begin{center}
\includegraphics[width=0.48\textwidth]{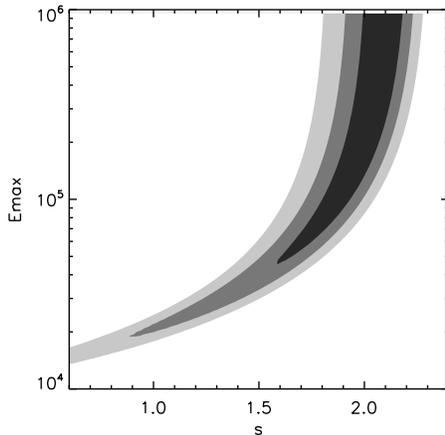}
\end{center}
\vspace*{-1.6cm}
\caption{Confidence contour as in Fig.~\ref{Fig:CLSigmaS} but for an exponential term $\exp(-E/E_{\textrm{max}})$ in Eq.~(\ref{EQ:SNRSpectra}) instead of the step function.}
\label{Fig:CLSigmaS.EXP.HE}
\end{figure}

Note the step function term $\Theta(E_{\textrm{max}}-E)$ in Eq.~(\ref{EQ:SNRSpectra}) can be replaced by an exponential term $\exp(-E/E_{\textrm{max}})$. 
Fig.~\ref{Fig:CLSigmaS.EXP.HE} shows the confidence level of $E_{\textrm{max}}$ vs. $s$ for the new best-fit parametrization. 
The contour shows a minor deviation from original confidence contours in Fig.~\ref{Fig:CLSigmaSEmax}, however, the conclusion is not altered.

\subsection{Conclusions}
We have considered a full picture of the hadronic $\gamma$-rays in cosmic-ray interactions and introduced an easy-to-use $\gamma$-ray production matrix which can be 
used for arbitrary cosmic-ray spectrum. The matrices are available for download at website {\em http://cherenkov.physics.iastate.edu/gamma-prod}. 
We apply the production matrix to calculate the $\gamma$-ray GeV excess and also the TeV-band spectrum of SNR RX J1713.7-3946. We conclude that 1) the modifications in 
the GeV-band $\gamma$-ray emission of 
hadronic origin are insufficient to explain the GeV excess in diffuse galactic $\gamma$-rays; 2) a soft cut-off at about 100~TeV is statistically required in the 
particle spectrum if the TeV-band spectrum of RX J1713.7-3946 as observed with HESS is caused by cosmic-ray nucleons; 3) no evidence for efficient nucleon acceleration 
to energies near the knee in the cosmic-ray spectrum, nor evidence of the spectral curvature and hardness predicted by standard models of cosmic-ray modified shock 
acceleration. We emphasize the need for GLAST data to better constrain the $\gamma$-ray spectrum below 100~GeV. 

In addition to $\gamma$-ray production matrix, we have also produced the production matrices of 
all particles that are stable on a time scale relevant for cosmic-ray propagation. The leptonic production by the same approach will be discussed in an independent 
article.
%

%
\subsection{Acknowledgments}
Grant support from NASA with award No. NAG5-13559 is gratefully acknowledged.
\bibliography{libros}
\bibliographystyle{plain}

\end{document}